# Understanding Small-Signal Impedance Matrices in Different Reference Frames

J. Pedra

***Abstract*—This paper systematically analyzes the relationships among the *dq*-domain, *αβ*-domain, and sequence-domain representations used in small-signal impedance modeling of voltage-source converters (VSCs). It is shown that the AC impedance matrix expressed with *dq*-complex and *αβ*-complex variables leads to different formulations in the sequence domain. The study demonstrates that asymmetric systems exhibit different physical phenomena in the rotating and stationary reference frames; therefore, the transformations between these frames are not physically consistent in such cases. It is also demonstrated that the so-called modified sequence-domain impedance is equivalent to the universal impedance model in the frequency domain. The analysis clarifies several notational inconsistencies found in the literature. Finally, a physical interpretation is presented highlighting the implications of using stationary and rotating reference frames for stability analysis of power converters.**

***Index Terms*—Complex and real transformations, symmetrical components, small-signal impedance, voltage-source converters.**

## I. Introduction

The application of time-dependent transformations, such as the Park transformation, has been largely restricted to electrical machine theory and control during the past century [1]–[3]. In recent decades, however, stability studies of VSC-HVDC systems in the *dq* and sequence domains have become a widely used analysis method. Both state-space and frequency-domain approaches are commonly applied to assess instability phenomena in these systems [4]. Frequency-domain methods based on system impedance characterization are particularly attractive because they can be applied using both analytical models and experimental measurements. These approaches originate from techniques originally developed for DC systems [5]. Impedance modeling of three-phase VSC-HVDC systems is normally performed in the *dq*-real domain [6] or in the sequence domain [7]–[9], resulting in a 2 × 2 impedance matrix. The off-diagonal terms of the impedance matrix describe the positive- and negative-sequence coupling. This effect is not considered in [7]. A comprehensive description of the *dq*-complex formulation can be found in [10], [11]. The relationship between the impedance matrix $Z_{dq}$ in the *dq*-real frame and the sequence-domain matrix $Z_{pn}$ is presented in [9], giving rise to the so-called modified sequence-domain impedance (MSDI), which incorporates the mirror frequency effect (MFE). A rigorous derivation based on the *dq*-complex frame is provided in [12]. Reference [8] shows that the sequence-domain impedance $Z_{pn}$ has the same marginal stability conditions as the *dq*-domain impedance matrix $Z_{dq}$. The transformation from the *αβ*–real frame to the *dq*–real frame is presented in [13]. This transformation is nontrivial because it involves transitioning from a stationary reference frame to a rotating reference frame. This results from the application of the Park transformation to the voltage and current variables. Examples of such frame transformations can be found in [14], [15]. Transformations from the *dq*-complex domain to the *αβ*-complex domain are described in [16], [17]. These voltage–current relationships are referred to as the unified impedance model (UIM) and are claimed to be formulated in a stationary reference frame. However, as demostrated in Section V-C, the UIM is equivalent to the MSDI, consequently, both are formulated in a rotating reference frame. An additional case of interest, described in [18], [19], includes the DC side in the analysis, resulting in a 3 × 3 impedance/admittance matrix. A global view of the relationships among different frames has been presented in [3] and [20], however, these works contain several issues that will be discussed later. The notation associated with *dq* reference frames (*dq*-real and *dq*-frequency in Fig. 1) is well standardized throughout the literature. However, no clear consensus has been reached regarding the notation for *pn* reference frames (*dq*-complex and *pn*-rotating in Fig. 1),

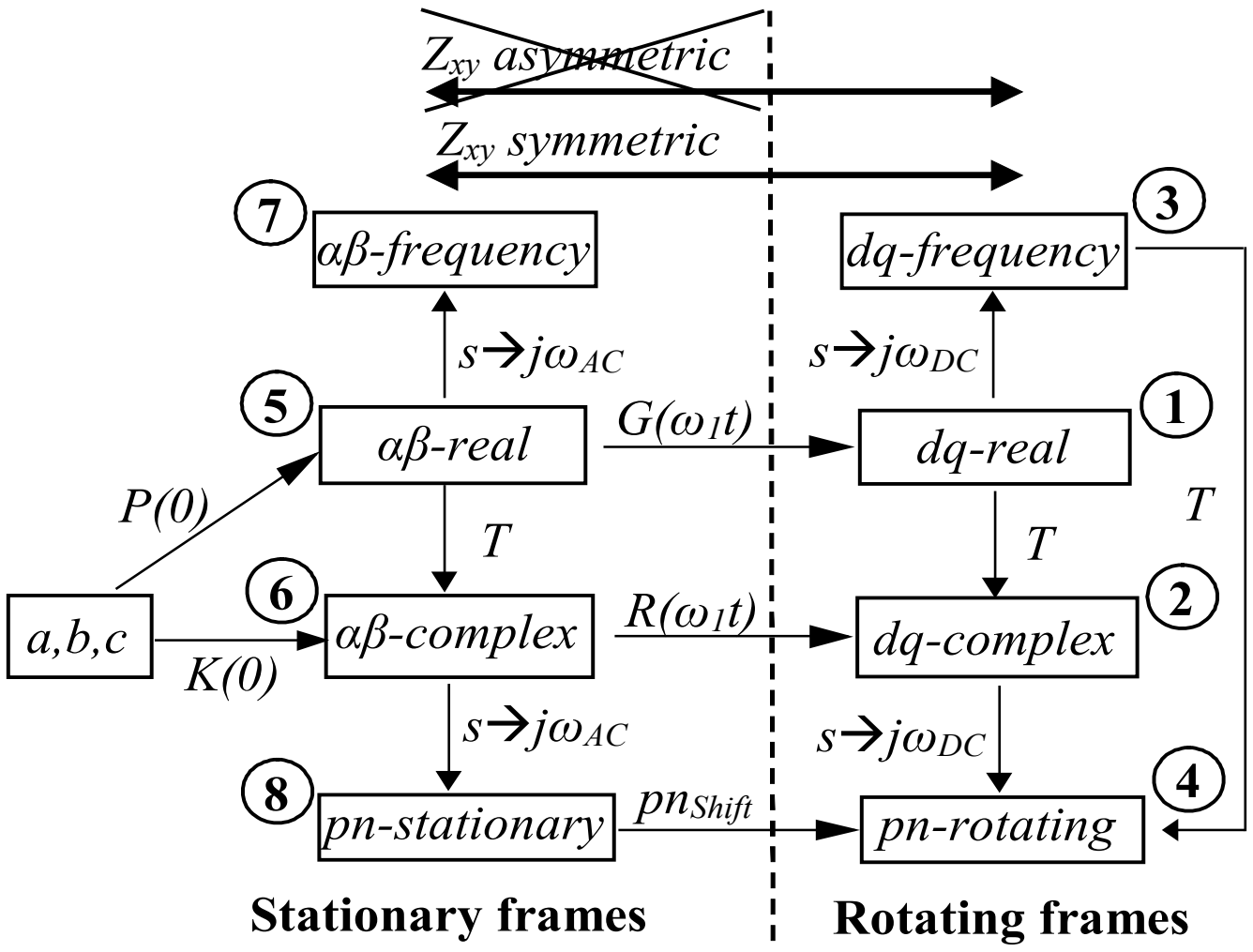

Fig. 1. Overview of transformations.

J. Pedra is with the Dep. of Elect. Eng., UPC, Av. Diagonal 647, 08028 Barcelona, Spain (mail: joaquin.pedra@upc.edu).

resulting in inconsistencies among references [20]–[29], as discussed in Section VII. Fig. 1 illustrates the relationships among the different reference frames and highlights the distinction between stationary and rotating frames. This distinction is important because transformations between these frames are physically meaningful only when the impedance matrix $Z_{xy}$ is symmetric. This impossibility of transforming from a stationary frame to a rotating frame in the presence of asymmetric impedance or admittance matrices is indicated in Fig. 1 by a dashed line and crossing out $Z_{xy}$. This result constitutes one of the main contributions of this paper. As discussed later, several studies, such as [21] and [22], attempt to perform a reference-frame transformation on asymmetric matrices, despite the questionable physical interpretation of such an operation. Fig. 1 also shows that the natural angular frequency in a stationary frame is $\omega_{AC}$. In contrast, for rotating frames, the natural angular frequency is $\omega_{DC} = \omega_{AC} - \omega_1$, where $\omega_1 = 2\pi f_1$, and $f_1$ is the fundamental frequency. The modified sequence-domain impedance [9] and the universal impedance model [16] correspond to position (**4**), *pn*-rotating, as shown in Fig. 1.

The main contributions of this paper are as follows:

1) A comprehensive diagram is provided in Fig. 1 to illustrate the transformations among the different reference frames and to clarify their relationships. The figure specifically emphasizes the existence of two distinct *pn* frames: a stationary *pn* frame and a rotating *pn* frame.
2) A physical justification is provided (Section VI) to demostrate why asymmetric impedance/admittance matrices cannot be transformed from a stationary frame to a rotating frame, or vice versa.
3) A demonstration is provided to establish the equivalence between the modified sequence-domain impedance and the universal impedance model in the frequency domain.
4) A physical explanation is provided for the frequency-coupling phenomenon that gives rise to two coupled frequencies in a rotating reference frame (Appendix III).
5) A standard notation framework for the sequence domain is proposed for analysis in both the Laplace and frequency domains.

## II. Relationship Between Transformations

The Park (*dq*-real) and Ku (*dq*-complex) transformations [1], [2] are reviewed in this section. Fig. 1 provides an overview of the transformations, excluding the homopolar component because VSCs do not include a neutral conductor. This overview is similar to those presented in [3] and [22], but it is extended here to include the relationships among the *pn*-sequence, *dq*-complex, and *αβ*-complex domains. Throughout this paper, normalized transformations are used. The case of non-normalized transformations is discussed in Appendix I.

The normalized Park (*dq*-real) transformation [10] of a three-phase voltage vector $U_{abc} = [U_a\ U_b\ U_c]^{\mathrm{T}}$ is defined as $U_{dq} = P(\theta)\ U_{abc}$, where

$$P(\theta) = \sqrt{\frac{2}{3}}\begin{bmatrix} \cos(\theta) & \cos(\theta - 2\pi/3) & \cos(\theta + 2\pi/3) \\ -\sin(\theta) & -\sin(\theta - 2\pi/3) & -\sin(\theta + 2\pi/3) \end{bmatrix}, \quad (1)$$

where $U_{dq} = [U_d\ \ U_q]^{\mathrm{T}}$ is the *dq*-real voltage vector. The voltage and current vectors, $I_{dq} = [I_d\ I_q]^{\mathrm{T}}$, are related through the impedance matrix $U_{dq} = Z_{dq}(s)I_{dq}$. The transformation angle $\theta$ is obtained from the PLL, or $\theta = \omega_1 t$ when the PLL is not considered, where $\omega_1 = 2\pi f_1$ and $f_1$ is the fundamental frequency. The inverse Park transformation is $U_{abc} = P(\theta)^{\mathrm{T}} U_{dq}$.

The *αβ*-real voltage vector $U_{\alpha\beta} = [U_\alpha\ U_\beta]^{\mathrm{T}}$ can be obtained from the three-phase set of voltages $U_{abc}$ by applying the Park (*dq*-real) transformation with $\theta = 0$, i.e., $U_{\alpha\beta} = P(0)U_{abc}$. The relation between the two transformations, $P(\theta) = G(\theta)\ P(0)$, is

$$G(\theta) = \begin{bmatrix} \cos(\theta) & \sin(\theta) \\ -\sin(\theta) & \cos(\theta) \end{bmatrix}. \quad (2)$$

The normalized Ku (*dq*-complex) transformation [10], [11] of a three-phase voltage vector $\mathbf{U^m_{dq}} = [\mathbf{U_{dq}}\ \ \mathbf{U_{dq}}^*]^{\mathrm{T}}$ is $\mathbf{U^m_{dq}} = K(\theta)\ U_{abc}$, where

$$K(\theta) = \frac{1}{\sqrt{3}}\begin{bmatrix} e^{-j\theta} & a \cdot e^{-j\theta} & a^2 \cdot e^{-j\theta} \\ e^{j\theta} & a^2 \cdot e^{j\theta} & a \cdot e^{j\theta} \end{bmatrix}, \quad (3)$$

and $a = e^{j\cdot 2\pi/3}$, $U_{abc} = (K(\theta)^{\mathrm{T}})^*\ \mathbf{U^m_{dq}}$, and $\mathbf{U_{dq}} = (U_d + j\cdot U_q)/\sqrt{2}$ is the *dq*-complex space voltage vector. This vector and its conjugate are also the forward and backward *dq*-complex space vectors (i.e., $\mathbf{U_f} = \mathbf{U_{dq}}$ and $\mathbf{U_b} = \mathbf{U^*_{dq}}$) in the literature on electrical machines [1].

The relation between the Park (*dq*-real) and Ku (*dq*-complex) transformations is $\mathbf{U^m_{dq}} = T\cdot U_{dq}$ and $U_{dq} = T^{-1}\cdot \mathbf{U^m_{dq}}$, i.e., $K(\theta) = T\cdot P(\theta)$, where

$$T = \frac{1}{\sqrt{2}}\begin{bmatrix} 1 & j \\ 1 & -j \end{bmatrix}, \cdot \quad (4)$$

and $T^{-1} = T^{\mathrm{H}}$, where superscript H denotes transpose and complex conjugate (i.e., Hermitian conjugate). The matrix $T$ also relates the *αβ*-real and *αβ*-complex domains.

The *αβ*-complex voltage vector $\mathbf{U^m_{\alpha\beta}} = \begin{bmatrix} \mathbf{U_{\alpha\beta}} & \mathbf{U^*_{\alpha\beta}} \end{bmatrix}^T$, where $\mathbf{U_{\alpha\beta}} = (U_\alpha + j\cdot U_\beta)/\sqrt{2}$, can be obtained from the three-phase set of voltages $U_{abc}$ by applying the Ku (*dq*-complex) transformation with $\theta = 0$, $\mathbf{U^m_{\alpha\beta}} = K(0)\cdot U_{abc}$. The relation between the two transformations is $K(\theta) = R(\theta)\ K(0)$, where

$$R(\theta) = \begin{bmatrix} e^{-j\theta} & 0 \\ 0 & e^{j\theta} \end{bmatrix}. \quad (5)$$

The *pn*-sequence emerges when the Laplace operator $s$ is replaced by $j\omega$ (see Fig. 1). The *pn*-sequence variables can be obtained from the *αβ*-complex variables by replacing the operator $s$ by $j\omega_{AC}$, or from the *dq*-complex variables by replacing $s$ by $j\omega_{DC}$. The angular frequencies $\omega_{AC} = 2\pi f_{AC}$ and $\omega_{DC} = 2\pi f_{DC}$ are the angular frequencies of the *αβ*- and *dq*-complex domains, respectively, where $\omega_{AC} = \omega_{DC} + \omega_1$. The frequencies $f_{AC} = \omega_{AC}/(2\pi)$, $f_{DC} = \omega_{DC}/(2\pi)$ are the AC and DC frequencies, respectively.

## III. Relationships Between Real and Complex Domains

This section describes the relationships among the $dq$-real, $dq$-complex, $\alpha\beta$-real, and $\alpha\beta$-complex domains.

### A. Relation Between dq-Real and dq-Complex Domains

According to [6], [8] and [9], a generic $dq$-real-domain impedance matrix can be written as

$$\begin{bmatrix} U_d(s) \\ U_q(s) \end{bmatrix} = \begin{bmatrix} Z_{dd}(s) & Z_{dq}(s) \\ Z_{qd}(s) & Z_{qq}(s) \end{bmatrix} \begin{bmatrix} I_d(s) \\ I_q(s) \end{bmatrix}, \quad (6)$$

and, using the admittance matrix, this relation becomes

$$\begin{bmatrix} I_d(s) \\ I_q(s) \end{bmatrix} = \begin{bmatrix} Y_{dd}(s) & Y_{dq}(s) \\ Y_{qd}(s) & Y_{qq}(s) \end{bmatrix} \begin{bmatrix} U_d(s) \\ U_q(s) \end{bmatrix}. \quad (7)$$

The impedance matrix in the $dq$-real frame can be transformed into a $dq$-complex-domain impedance matrix by applying the following relations

$$\begin{bmatrix} \mathbf{U_{dq}} \\ \mathbf{U^*_{dq}} \end{bmatrix} = \frac{1}{\sqrt{2}} \begin{bmatrix} 1 & j \\ 1 & -j \end{bmatrix} \begin{bmatrix} U_d \\ U_q \end{bmatrix} \quad ; \quad \begin{bmatrix} U_d \\ U_q \end{bmatrix} = \frac{1}{\sqrt{2}} \begin{bmatrix} 1 & 1 \\ -j & j \end{bmatrix} \begin{bmatrix} \mathbf{U_{dq}} \\ \mathbf{U^*_{dq}} \end{bmatrix}, \quad (8)$$

which results in

$$\begin{gathered} \begin{bmatrix} \mathbf{U_{dq}} \\ \mathbf{U^*_{dq}} \end{bmatrix} = \frac{1}{\sqrt{2}} \begin{bmatrix} 1 & j \\ 1 & -j \end{bmatrix} \begin{bmatrix} Z_{dd}(s) & Z_{dq}(s) \\ Z_{qd}(s) & Z_{qq}(s) \end{bmatrix} \frac{1}{\sqrt{2}} \begin{bmatrix} 1 & 1 \\ -j & j \end{bmatrix} \begin{bmatrix} \mathbf{I_{dq}} \\ \mathbf{I^*_{dq}} \end{bmatrix} \\ \begin{bmatrix} \mathbf{U_{dq}}(s) \\ \mathbf{U^*_{dq}}(s) \end{bmatrix} = \begin{bmatrix} \mathbf{Z}^+_{dq}(s) & \mathbf{Z}^-_{dq}(s) \\ (\mathbf{Z}^-_{dq})^*(s) & (\mathbf{Z}^+_{dq})^*(s) \end{bmatrix} \begin{bmatrix} \mathbf{I_{dq}}(s) \\ \mathbf{I^*_{dq}}(s) \end{bmatrix}, \end{gathered} \quad (9)$$

where

$$\begin{aligned} \mathbf{Z}^+_{dq}(s) &= \frac{Z_{dd}(s)+Z_{qq}(s)}{2} + j\frac{Z_{qd}(s)-Z_{dq}(s)}{2} \\ \mathbf{Z}^-_{dq}(s) &= \frac{Z_{dd}(s)-Z_{qq}(s)}{2} + j\frac{Z_{qd}(s)+Z_{dq}(s)}{2}. \end{aligned} \quad (10)$$

Eq. (6) is expressed in Laplace domain, therefore, $U_d$, $U_q$, $I_d$, and $I_q$ are real functions of the variable $s$. Eq. (6) corresponds to the box **(1)**, titled $dq$-real, in Fig. 1. Similarly, Eq. (9) corresponds to the box **(2)** titled $dq$-complex in Fig. 1. Similar relations to (10) can be obtained for the admittance matrix. The inverse relation between $dq$-real and $dq$-complex impedance matrices is also of interest and is defined as

$$\begin{aligned} Z_{dd}(s) &= \frac{1}{2}\left[ \mathbf{Z}^+_{dq}(s) + \mathbf{Z}^-_{dq}(s) + \left(\mathbf{Z}^+_{dq}(s) + \mathbf{Z}^-_{dq}(s)\right)^* \right] \\ Z_{qq}(s) &= \frac{1}{2}\left[ \mathbf{Z}^+_{dq}(s) - \mathbf{Z}^-_{dq}(s) + \left(\mathbf{Z}^+_{dq}(s) - \mathbf{Z}^-_{dq}(s)\right)^* \right] \\ Z_{dq}(s) &= \frac{1}{2}\left[ j\left(\mathbf{Z}^+_{dq}(s) - \mathbf{Z}^-_{dq}(s)\right) - j\left(\mathbf{Z}^+_{dq}(s) - \mathbf{Z}^-_{dq}(s)\right)^* \right] \\ Z_{qd}(s) &= \frac{1}{2}\left[ -j\left(\mathbf{Z}^+_{dq}(s) + \mathbf{Z}^-_{dq}(s)\right) + j\left(\mathbf{Z}^+_{dq}(s) + \mathbf{Z}^-_{dq}(s)\right)^* \right]. \end{aligned} \quad (11)$$

Note that the variable $s$ must be treated as real when the complex functions in the $s$-domain are conjugated in (9) and (11), i.e., the complex conjugate function of $Z(s) = A(s) + jB(s)$ becomes $Z^*(s) = A(s) - jB(s)$ [11].

### B. Relation Between αβ-Real and αβ-Complex Domains

The $\alpha\beta$ domain corresponds to $\theta = 0$ in the Park and Ku transformations. The relationship between voltages, $U_\alpha$, $U_\beta$, and currents $I_\alpha$, $I_\beta$, which are real functions of the variable $s$, is

$$\begin{bmatrix} U_\alpha(s) \\ U_\beta(s) \end{bmatrix} = \begin{bmatrix} Z_{\alpha\alpha}(s) & Z_{\alpha\beta}(s) \\ Z_{\beta\alpha}(s) & Z_{\beta\beta}(s) \end{bmatrix} \begin{bmatrix} I_\alpha(s) \\ I_\beta(s) \end{bmatrix}. \quad (12)$$

Eq. (12) corresponds to the box **(5)**, labeled $\alpha\beta$-real in Fig. 1. It should be noted that the reference frame is now stationary, unlike the previous frames, which are rotating. The relationship between the variables in the $\alpha\beta$-real and $\alpha\beta$-complex domains is

$$\begin{bmatrix} \mathbf{U_{\alpha\beta}} \\ \mathbf{U^*_{\alpha\beta}} \end{bmatrix} = \frac{1}{\sqrt{2}} \begin{bmatrix} 1 & j \\ 1 & -j \end{bmatrix} \begin{bmatrix} U_\alpha \\ U_\beta \end{bmatrix} \quad ; \quad \begin{bmatrix} U_\alpha \\ U_\beta \end{bmatrix} = \frac{1}{\sqrt{2}} \begin{bmatrix} 1 & 1 \\ -j & j \end{bmatrix} \begin{bmatrix} \mathbf{U_{\alpha\beta}} \\ \mathbf{U^*_{\alpha\beta}} \end{bmatrix}. \quad (13)$$

In the $\alpha\beta$-complex domain, the relationship between voltages, $\mathbf{U}_{\alpha\beta}$, $\mathbf{U}^*_{\alpha\beta}$, and currents $\mathbf{I}_{\alpha\beta}$ and $\mathbf{I}^*_{\alpha\beta}$, which are complex functions of the variable $s$, is

$$\begin{bmatrix} \mathbf{U_{\alpha\beta}}(s) \\ \mathbf{U^*_{\alpha\beta}}(s) \end{bmatrix} = \begin{bmatrix} \mathbf{Z}^+_{\alpha\beta}(s) & \mathbf{Z}^-_{\alpha\beta}(s) \\ \left(\mathbf{Z}^-_{\alpha\beta}\right)^*(s) & \left(\mathbf{Z}^+_{\alpha\beta}\right)^*(s) \end{bmatrix} \begin{bmatrix} \mathbf{I_{\alpha\beta}}(s) \\ \mathbf{I^*_{\alpha\beta}}(s) \end{bmatrix}. \quad (14)$$

Eq. (14) corresponds to box **(6)**, labeled $\alpha\beta$-complex in Fig. 1. The relationship between the elements of the impedance matrix in $\alpha\beta$-real and $\alpha\beta$-complex frames is

$$\begin{aligned} \mathbf{Z}^+_{\alpha\beta}(s) &= \frac{Z_{\alpha\alpha}(s)+Z_{\beta\beta}(s)}{2} + j\frac{Z_{\beta\alpha}(s)-Z_{\alpha\beta}(s)}{2} \\ \mathbf{Z}^-_{\alpha\beta}(s) &= \frac{Z_{\alpha\alpha}(s)-Z_{\beta\beta}(s)}{2} + j\frac{Z_{\beta\alpha}(s)+Z_{\alpha\beta}(s)}{2}. \end{aligned} \quad (15)$$

### C. Unified impedance model

The voltage–current relationships of the so-called "unified impedance model" introduced in [16] are

$$\begin{bmatrix} \mathbf{U_{\alpha\beta}} \\ e^{j2\omega_1 t}\mathbf{U^*_{\alpha\beta}} \end{bmatrix} = \begin{bmatrix} \mathbf{Z}^+_{dq}(s-j\omega_1) & \mathbf{Z}^-_{dq}(s-j\omega_1) \\ (\mathbf{Z}^-_{dq})^*(s-j\omega_1) & (\mathbf{Z}^+_{dq})^*(s-j\omega_1) \end{bmatrix} \begin{bmatrix} \mathbf{I_{\alpha\beta}} \\ e^{j2\omega_1 t}\mathbf{I^*_{\alpha\beta}} \end{bmatrix}. \quad (16)$$

In [16] and [17], this impedance model is presented as a stationary-frame model; however, this classification is incorrect. This issue is further discussed in Sections V-C and VI.

## IV. Relationship Between Static and Rotating Reference Frames

As shown in Fig. 1 and justified in Section VI, transformations between stationary and rotating frames can only be applied to symmetric impedance or admittance matrices. An impedance/admittance matrix in the $dq$-real frame is symmetric if $Z_{dd}(s) = Z_{qq}(s)$ and $Z_{dq}(s) = -Z_{qd}(s)$ ($Y_{dd} = Y_{qq}$ and $Y_{dq} = -Y_{qd}$). In the $dq$-complex frame, symmetry implies the condition $\mathbf{Z}^{-}_{dq}(s) = 0$ ($\mathbf{Y}^{-}_{dq} = 0$). Similar conditions apply to the $\alpha\beta$-real and $\alpha\beta$-complex reference frames.

### A. Relation between αβ-Complex and dq-Complex Reference Frames

First, the case of frame transformation using symmetric complex impedance matrices is considered because it is simpler than the case involving real matrices. The superscripts $s$ and $r$ are used to indicate stationary and rotating reference frames, respectively. As explained in [10], transforming the relation $y^s = G(s)u^s$ in a stationary frame to a rotating frame results in the substitution $s \to s+j\omega_1$, that is, $y^r = G(s+j\omega_1)u^r$. Similarly, to change from a rotating frame to a stationary one, the translation $s \to s-j\omega_1$ can be applied [10].

As an example, a series impedance in a three-phase system has a symmetric impedance matrix in the $\alpha\beta$-complex reference frame

$$Z_{\alpha\beta}(s) = Z^{+}_{\alpha\beta}(s) = \begin{bmatrix} R+sL & 0 \\ 0 & R+sL \end{bmatrix}. \tag{17}$$

To change the frame from stationary to rotating, it must be taken into account that the matrix $\mathbf{Z}^{+}_{dq}$ has the term $\mathbf{Z}^{+}(s)$ and the complex-conjugate term, $(\mathbf{Z}^{+}(s))^{*}$. Thus, applying the substitution $s \to s+j\omega_1$ [10] results in

$$\mathbf{Z}^{+}_{dq}(s) = \begin{bmatrix} R+(s+j\omega_1)L & 0 \\ 0 & R+(s-j\omega_1)L \end{bmatrix}. \tag{18}$$

The frame change can also be performed using symmetric admittance matrices. Symmetry imposes that the impedance matrix is diagonal, which makes the inverse trivial. Applying this inverse to (18) yields

$$\mathbf{Y}^{+}_{dq}(s) = \begin{bmatrix} \dfrac{1}{R+(s+j\omega_1)L} & 0 \\ 0 & \dfrac{1}{R+(s-j\omega_1)L} \end{bmatrix}. \tag{19}$$

### B. Relation Between αβ-Real and dq-Real Reference Frames

The transformation from the $\alpha\beta$-real reference frame to the $dq$-real reference frame [13] is more complicated than the transformation from $\alpha\beta$-complex frame to $dq$-complex frame.

Reference [13] describes the transformation of a matrix $H^s(s)$ in the stationary reference frame into a matrix $H^r(s)$ in the rotating reference frame. Since the procedure described in [13] is difficult to follow, a more straightforward approach is to first transform from the $\alpha\beta$-real reference-frame to the $\alpha\beta$-complex representation, then obtain the $dq$-complex representation by applying the frequency shift $s \to s+j\omega_1$ , and finally convert it to the $dq$-real representation using (11). Applying this procedure to the series impedance in a three-phase system given by (17) where the frequency shift yields (18), and the application of (11) finally results in

$$Z_{dq} = \begin{bmatrix} R+sL & -\omega_1 L \\ \omega_1 L & R+sL \end{bmatrix}. \tag{20}$$

Reference [13] also describes the transformation from the matrix $H^r(s)$ in the rotating frame to the matrix $H^s(s)$ in the stationary frame. Similarly to the previous case, a more straightforward approach is to first transform from the $dq$-real reference frame to the $dq$-complex reference frame, then apply the frequency shift $s \to s-j\omega_1$ [10], to obtain the $\alpha\beta$-complex reference frame, and finally transform to the $\alpha\beta$-real reference frame using the $\alpha\beta$-frame version of (11). Eq. (21) shows the impedance matrix of a PI controller in a rotating frame, $Z_{dq}(s)$.

$$Z_{dq} = \begin{bmatrix} k_p + \dfrac{k_i}{s} & 0 \\ 0 & k_p + \dfrac{k_i}{s} \end{bmatrix}. \tag{21}$$

Using (10), we have $Z_{dq}(s) = \mathbf{Z}^{+}_{dq}(s)$. Applying the translation $s \to s-j\omega_1$ yields the expression for $\mathbf{Z}^{+}_{\alpha\beta}$ in (22)

$$\mathbf{Z}^{+}_{\alpha\beta} = \begin{bmatrix} k_p + \dfrac{k_i}{s-j\omega_1} & 0 \\ 0 & k_p + \dfrac{k_i}{s+j\omega_1} \end{bmatrix}, \tag{22}$$

and, using the version of (11) for $\alpha\beta$-frames, (23) is obtained.

$$Z_{\alpha\beta} = \begin{bmatrix} k_p + \dfrac{k_i s}{s^2+\omega_1^2} & \dfrac{-k_i\omega_1}{s^2+\omega_1^2} \\ \dfrac{k_i\omega_1}{s^2+\omega_1^2} & k_p + \dfrac{k_i s}{s^2+\omega_1^2} \end{bmatrix}. \tag{23}$$

## V. Frequency-Domain Representations

After relating the $dq$-real and $dq$-complex domains, these variables must be connected to the positive- and negative-sequence components of a three-phase system. The system relates the positive-sequence component at $\omega_{AC} = \omega_{DC} + \omega_1$ with a negative-sequence component at $\omega_{AC} = \omega_{DC} - \omega_1$. Applying the Ku transformation to both components produces the angular frequency $\omega_{DC}$. Up to this point, the analysis has used real or complex Laplace-domain variables. From this point onward, phasor notation is adopted, indicated by an underline. The resulting voltages and currents in the $dq$-complex frame are

$$\begin{aligned} \mathbf{U}_{\mathrm{dq}} &= \underline{U}_p e^{j\omega_{DC}t} + \underline{U}_n^{*} e^{-j\omega_{DC}t} \; ; \mathbf{I}_{\mathrm{dq}} = \underline{I}_p e^{j\omega_{DC}t} + \underline{I}_n^{*} e^{-j\omega_{DC}t} \\ \mathbf{U}^{*}_{\mathrm{dq}} &= \underline{U}_p^{*} e^{-j\omega_{DC}t} + \underline{U}_n e^{j\omega_{DC}t} \; ; \mathbf{I}^{*}_{\mathrm{dq}} = \underline{I}_p^{*} e^{-j\omega_{DC}t} + \underline{I}_n e^{j\omega_{DC}t}. \end{aligned} \tag{24}$$

These equations are derived in Appendices II and III. The phasors $\underline{U}_p$, $\underline{U}_n$, $\underline{I}_p$, y $\underline{I}_n$ are related to the positive- and negative-sequence components of the voltages and currents, respectively, up to a scaling factor, which is omitted for notational simplicity. Using the operator $ph_{\omega}(\mathbf{X})$, which extracts the phasor of time-domain variable $\mathbf{X}$ at the angular frequency $\omega$, the phasors associated with $\mathbf{U}_{\mathrm{dq}}$, $\mathbf{U}^{*}_{\mathrm{dq}}$, $\mathbf{I}_{\mathrm{dq}}$, and $\mathbf{I}^{*}_{\mathrm{dq}}$ at the angular frequency $+\omega_{DC}$, are given by

$$ph_{\omega}\left(\mathbf{U_{dq}}\right)=\underline{U}_p=U_p\angle\alpha_p \quad ; \quad ph_{\omega}\left(\mathbf{U}^*_{dq}\right)=\underline{U}_n=U_n\angle\alpha_n$$
$$ph_{\omega}\left(\mathbf{I_{dq}}\right)=\underline{I}_{\mathbf{p}}=I_p\angle\beta_p \quad ; \quad ph_{\omega}\left(\mathbf{I}^*_{dq}\right)=\underline{I}_{\mathbf{n}}=I_n\angle\beta_n. \quad (25)$$

Since the voltages and currents in (24) are time-dependent, the Laplace operator $s$ is defined as $s = d/dt$. From these expressions, the $dq$-real variables are

$$\begin{aligned}\mathbf{U}_d &=\left|\underline{U}_p\right|\cos\left(\omega_{DC}t+\alpha_p\right)+\left|\underline{U}_n\right|\cos\left(\omega_{DC}t+\alpha_n\right)\\ \mathbf{U}_q &=\left|\underline{U}_p\right|\mathrm{sen}\left(\omega_{DC}t+\alpha_p\right)-\left|\underline{U}_n\right|\mathrm{sen}\left(\omega_{DC}t+\alpha_n\right)\\ \mathbf{I}_d &=\left|\underline{I}_p\right|\cos\left(\omega_{DC}t+\beta_p\right)+\left|\underline{I}_n\right|\cos\left(\omega_{DC}t+\beta_n\right)\\ \mathbf{I}_q &=\left|\underline{I}_p\right|\mathrm{sen}\left(\omega_{DC}t+\beta_p\right)-\left|\underline{I}_n\right|\mathrm{sen}\left(\omega_{DC}t+\beta_n\right).\end{aligned} \quad (26)$$

By applying (8), defined in the Laplace domain, the following relationships in the frequency domain are obtained between the $dq$-axis voltage and current phasors and the positive- and negative-sequence voltage and current phasors at the angular frequency $+\omega_{DC}$:

$$\underline{U}_d=\frac{\underline{U}_p+\underline{U}_n}{\sqrt{2}}=U_d\angle\alpha_d \quad ; \quad \underline{U}_q=\frac{-j\underline{U}_p+j\underline{U}_n}{\sqrt{2}}=U_q\angle\alpha_q$$
$$\underline{I}_d=\frac{\underline{I}_p+\underline{I}_n}{\sqrt{2}}=I_d\angle\beta_d \quad ; \quad \underline{I}_q=\frac{-j\underline{I}_p+j\underline{I}_n}{\sqrt{2}}=I_q\angle\beta_q, \quad (27)$$

leading to the following properties:

- $\underline{U}_d$ is purely direct-axis ($\underline{U}_q = 0$) when $\underline{U}_p = \underline{U}_n$.
- $\underline{U}_q$ is purely quadrature-axis ($\underline{U}_d = 0$) when $\underline{U}_p = -\underline{U}_n$.

The inverse relation of (27) is

$$\underline{U}_p=\frac{\underline{U}_d+j\underline{U}_q}{\sqrt{2}}=U_p\angle\alpha_p \quad ; \quad \underline{U}_n=\frac{\underline{U}_d-j\underline{U}_q}{\sqrt{2}}=U_n\angle\alpha_n. \quad (28)$$

*A. dq-Frequency and pn-Rotating Frames*

Frame **(3)** in Fig. 1 is defined at $\omega_{DC}$; therefore $\underline{U}_d$, $\underline{U}_q$, $\underline{I}_d$, and $\underline{I}_q$ are phasors at $\omega_{DC}$.

$$\begin{bmatrix}\underline{U}_d(\omega_{DC})\\ \underline{U}_q(\omega_{DC})\end{bmatrix}=\begin{bmatrix}Z_{dd}(\omega_{DC}) & Z_{dq}(\omega_{DC})\\ Z_{qd}(\omega_{DC}) & Z_{qq}(\omega_{DC})\end{bmatrix}\begin{bmatrix}\underline{I}_d(\omega_{DC})\\ \underline{I}_q(\omega_{DC})\end{bmatrix}. \quad (29)$$

This $dq$-frequency frame is widely used [4], [6].

Frame **(4)** in Fig. 1 is also defined in the frequency domain, but with $\underline{U}_p$, $\underline{U}_n$, $\underline{I}_p$, and $\underline{I}_n$ at $\omega_{DC}+\omega_1$ and $\omega_{DC}-\omega_1$.

$$\begin{bmatrix}\underline{U}_p(\omega_{DC}+\omega_1)\\ \underline{U}_n(\omega_{DC}-\omega_1)\end{bmatrix}=\begin{bmatrix}Z^r_{pp}(\omega_{DC}) & Z^r_{pn}(\omega_{DC})\\ Z^r_{np}(\omega_{DC}) & Z^r_{nn}(\omega_{DC})\end{bmatrix}\begin{bmatrix}\underline{I}_p(\omega_{DC}+\omega_1)\\ \underline{I}_n(\omega_{DC}-\omega_1)\end{bmatrix}. \quad (30)$$

The impedance and admittance matrix $\boldsymbol{Z_{pn}}$ and $\boldsymbol{Y_{pn}}$ include the superscript $r$ to indicate that they correspond to a rotating frame. References [19] and [30] use this representation.

*B. αβ-Frequency and pn-Stationary Frames*

In the $\alpha\beta$-complex frame, using Appendix II with $P(\theta)$, where $\theta = \omega_1 t = 0$, yields

$$\mathbf{U_{\alpha\beta}}=\underline{U}_p e^{j\omega_{AC}t}+\underline{U}^*_n e^{-j\omega_{AC}t} \; ; \; \mathbf{I_{\alpha\beta}}=\underline{I}_p e^{j\omega_{AC}t}+\underline{I}^*_n e^{-j\omega_{AC}t}, \quad (31)$$

and in the $\alpha\beta$-real frame

$$\begin{aligned}\mathbf{U}_{\alpha} &=\left|\underline{U}_p\right|\cos\left(\omega_{AC}t+\phi_p\right)+\left|\underline{U}_n\right|\cos\left(\omega_{AC}t+\phi_n\right)\\ \mathbf{U}_{\beta} &=\left|\underline{U}_p\right|\mathrm{sen}\left(\omega_{AC}t+\phi_p\right)-\left|\underline{U}_n\right|\mathrm{sen}\left(\omega_{AC}t+\phi_n\right).\end{aligned} \quad (32)$$

Frame **(7)** in Fig. 1 ($\alpha\beta$-frequency) uses phasors at $\omega_{AC}$,

$$\begin{bmatrix}\underline{U}_\alpha(\omega_{AC})\\ \underline{U}_\beta(\omega_{AC})\end{bmatrix}=\begin{bmatrix}Z_{\alpha\alpha}(\omega_{AC}) & Z_{\alpha\beta}(\omega_{AC})\\ Z_{\beta\alpha}(\omega_{AC}) & Z_{\beta\beta}(\omega_{AC})\end{bmatrix}\begin{bmatrix}\underline{I}_\alpha(\omega_{AC})\\ \underline{I}_\beta(\omega_{AC})\end{bmatrix}. \quad (33)$$

Frame **(8)** ($pn$-stationary) also uses $\omega_{AC}$,

$$\begin{bmatrix}\underline{U}_p(\omega_{AC})\\ \underline{U}_n(\omega_{AC})\end{bmatrix}=\begin{bmatrix}Z^s_{pp}(\omega_{AC}) & Z^s_{pn}(\omega_{AC})\\ Z^s_{np}(\omega_{AC}) & Z^s_{nn}(\omega_{AC})\end{bmatrix}\begin{bmatrix}\underline{I}_p(\omega_{AC})\\ \underline{I}_n(\omega_{AC})\end{bmatrix}. \quad (34)$$

The superscript $s$ indicates that the impedance matrix is expressed in the stationary frame.

*C. Unified impedance model in the frequency domain..*

Rewriting the equations (31) to emphasize their dependence on the angular frequency $\omega_{AC}$, we obtain:

$$\begin{aligned}\mathbf{U_{\alpha\beta}} &=\underline{U}_p(\omega_{AC})e^{j\omega_{AC}t}+\underline{U}^*_n(\omega_{AC})e^{-j\omega_{AC}t}\\ \mathbf{I_{\alpha\beta}} &=\underline{I}_p(\omega_{AC})e^{j\omega_{AC}t}+\underline{I}^*_n(\omega_{AC})e^{-j\omega_{AC}t},\end{aligned} \quad (35)$$

consequently,

$$\begin{aligned}e^{j2\omega_1 t}\mathbf{U}^*_{\alpha\beta} &=\underline{U}_n(\omega_{AC})e^{j(\omega_{AC}+2\omega_1)t}+\underline{U}^*_p(\omega_{AC})e^{-j(\omega_{AC}-2\omega_1)t}\\ e^{j2\omega_1 t}\mathbf{I}^*_{\alpha\beta} &=\underline{I}_n(\omega_{AC})e^{j(\omega_{AC}+2\omega_1)t}+\underline{I}^*_p(\omega_{AC})e^{-j(\omega_{AC}-2\omega_1)t},\end{aligned} \quad (36))$$

where the phasors associated with $\mathbf{U}_{\alpha\beta}$, $\mathbf{U}^*_{\alpha\beta}$, $\mathbf{I}_{\alpha\beta}$, and $\mathbf{I}^*_{\alpha\beta}$ at an angular frequency $+\omega_x$, are given by $\omega_x = +\omega_{AC}$ for the positive sequence and $\omega_x = +\omega_{AC} - 2\omega_1$ for the negative sequence, resulting in

$$\begin{aligned}ph_{\omega}\left(\mathbf{U_{\alpha\beta}}\right) &=\underline{U}_p(\omega_{AC})=U_{p,\omega_{AC}}\angle\alpha_{p,\omega_{AC}}\\ ph_{\omega}\left(\mathbf{I_{\alpha\beta}}\right) &=\underline{I}_{\mathbf{p}}(\omega_{AC})=I_{p,\omega_{AC}}\angle\beta_{p,\omega_{AC}}\\ ph_{\omega}\left(e^{j2\omega_1 t}\mathbf{U}^*_{\alpha\beta}\right) &=\underline{U}_n(\omega_{AC}-2\omega_1)=U_{n,\omega_{AC}-2\omega_1}\angle\alpha_{n,\omega_{AC}-2\omega_1}\\ ph_{\omega}\left(e^{j2\omega_1 t}\mathbf{I}^*_{\alpha\beta}\right) &=\underline{I}_n(\omega_{AC}-2\omega_1)=I_{n,\omega_{AC}-2\omega_1}\angle\beta_{n,\omega_{AC}-2\omega_1},\end{aligned} \quad (37)$$

and the voltaje-current relationships of (16) are given by

$$\begin{bmatrix}\underline{U}_p(s)\\ \underline{U}_n(s-j2\omega_1)\end{bmatrix}=\begin{bmatrix}Z_{pp}(s-j\omega_1) & Z_{pn}(s-j\omega_1)\\ Z_{np}(s-j\omega_1) & Z_{nn}(s-j\omega_1)\end{bmatrix}\begin{bmatrix}\underline{I}_p(s)\\ \underline{I}_n(s-j2\omega_1)\end{bmatrix}, \quad (38))$$

where $s$ must be substitutied by $j\omega_{AC}$.

## VI. Consequences of the Asymmetry of the Impedance/Admittance Matrix

This section analyzes the system from a physical perspective using either a stationary or a rotating reference frame. Fig. 2 presents a summary of the symmetry conditions in different reference frames. The *dq*-real and *αβ*-real reference frames, which are the most commonly used in the literature, are not considered in this section because their direct and inverse components are linear combinations of the positive- and negative-sequence components (see Eq. (27)), which makes their physical interpretation difficult.

Fig. 3 summarizes the properties of the different reference frames and highlights the main relationships between them.

The main properties to be highlighted are:

1) The modified sequence-domain impedance (Eq. (30)) and the universal impedance model (Eq. (38)) provide equivalent descriptions of the system, as they are related through the frequency relation $\omega_{DC} = \omega_{AC} - \omega_1$. A key physical consequence is that, in a rotating reference frame, an asymmetric system inherently introduces coupling between voltages and currents at different angular frequencies.
2) Frequency coupling does not appear in the voltage–current relationships expressed in a stationary reference frame (Eq. (34), Fig. 3). Since the $P(0)$ and $K(0)$ transformations are composed exclusively of constant coefficients, they preserve the frequencies of the transformed variables and therefore relate only voltages and currents at the same angular frequency.
3) The existence of frequency coupling is directly linked to the symmetry properties of the impedance or admittance matrix. For symmetric matrices, no frequency coupling arises in the rotating reference frame, allowing an equivalent representation in either the stationary or the rotating frame. However, for asymmetric matrices, frequency coupling becomes unavoidable, and an equivalent stationary-frame representation no longer exists. Eqs. (30) and (34) represent two different physical situations; therefore, asymmetric impedance/admittance matrices cannot be transformed between stationary and rotating reference frames. This distinction is not explicitly stated in Refs. [21] and [22], where the impedance matrix is transformed from a rotating to a stationary reference frame without clarifying that such a transformation is valid only under symmetric conditions.
4) It is important to note that, even when the system is symmetric, the notation in Eqs. (34) and (38) is different. For the universal impedance model, one obtains

$$\begin{bmatrix} \underline{U}_p(\omega_{AC}) \\ \underline{U}_n(\omega_{AC}-2\omega_1) \end{bmatrix} = \begin{bmatrix} Z^r_{pp}(\omega_{AC}-\omega_1) & 0 \\ 0 & Z^r_{nn}(\omega_{AC}-\omega_1) \end{bmatrix} \begin{bmatrix} \underline{I}_p(\omega_{AC}) \\ \underline{I}_n(\omega_{AC}-2\omega_1) \end{bmatrix}, \quad (39)$$

and in the stationary reference frame, it follows that

$$\begin{bmatrix} \underline{U}_p(\omega_{AC}) \\ \underline{U}_n(\omega_{AC}) \end{bmatrix} = \begin{bmatrix} Z^s_{pp}(\omega_{AC}) & 0 \\ 0 & Z^s_{nn}(\omega_{AC}) \end{bmatrix} \begin{bmatrix} \underline{I}_p(\omega_{AC}) \\ \underline{I}_n(\omega_{AC}) \end{bmatrix}. \quad (40)$$

In the rotating reference frame, we have $\underline{U}_p(\omega_{AC}) = Z^r_{pp}(\omega_{AC} - \omega_1)\underline{I}_p(\omega_{AC})$, whereas in the stationary reference frame $\underline{U}_p(\omega_{AC}) = Z^s_{pp}(\omega_{AC})\underline{I}_p(\omega_{AC})$. This notation issue leads to the $pn_{shift}$ shifting procedure introduced in Eq. (41), which converts the *pn*-stationary formulation in (**8**) into

**Symmetry conditions**

<u>*dq*-real frame</u>: $Z_{dd}(s) = Z_{qq}(s)$ and $Z_{dq}(s) = -Z_{qd}(s)$

<u>*αβ*-real frame</u>: $Z_{\alpha\alpha}(s) = Z_{\beta\beta}(s)$ and $Z_{\alpha\beta}(s) = -Z_{\beta\alpha}(s)$

<u>*dq*-complex frame</u>: $\mathbf{Z}^-_{dq}(s) = 0$ ($\mathbf{Y}^-_{dq}(s) = 0$)

<u>*αβ*-complex frame</u>: $\mathbf{Z}^-_{\alpha\beta}(s) = 0$ ($\mathbf{Y}^-_{\alpha\beta}(s) = 0$)

<u>*pn*-rotating frame</u>: $Z^r_{pn}(\omega_{DC}) = Z^r_{np}(\omega_{DC}) = 0$

<u>*pn*-stationary frame</u>: $Z^s_{pn}(\omega_{AC}) = Z^s_{np}(\omega_{AC}) = 0$

Fig. 2. Summary of symmetry conditions.

**Rotating reference frame**

*Modified sequence-domaine impedance* (*pn-rotating*):

$$\begin{bmatrix} \underline{U}_p(\omega_{DC}+\omega_1) \\ \underline{U}_n(\omega_{DC}-\omega_1) \end{bmatrix} = \begin{bmatrix} Z^r_{pp}(\omega_{DC}) & Z^r_{pn}(\omega_{DC}) \\ Z^r_{np}(\omega_{DC}) & Z^r_{nn}(\omega_{DC}) \end{bmatrix} \begin{bmatrix} \underline{I}_p(\omega_{DC}+\omega_1) \\ \underline{I}_n(\omega_{DC}-\omega_1) \end{bmatrix}, \text{ Equation (30)}$$

*Universal impedance model*:

$$\begin{bmatrix} \underline{U}_p(\omega_{AC}) \\ \underline{U}_n(\omega_{AC}-2\omega_1) \end{bmatrix} = \begin{bmatrix} Z^r_{pp}(\omega_{AC}-\omega_1) & Z^r_{pn}(\omega_{AC}-\omega_1) \\ Z^r_{np}(\omega_{AC}-\omega_1) & Z^r_{nn}(\omega_{AC}-\omega_1) \end{bmatrix} \begin{bmatrix} \underline{I}_p(\omega_{AC}) \\ \underline{I}_n(\omega_{AC}-2\omega_1) \end{bmatrix}, \text{ Equation (38)}$$

**Stationary reference frame**

$$\begin{bmatrix} \underline{U}_p(\omega_{AC}) \\ \underline{U}_n(\omega_{AC}) \end{bmatrix} = \begin{bmatrix} Z^s_{pp}(\omega_{AC}) & Z^s_{pn}(\omega_{AC}) \\ Z^s_{np}(\omega_{AC}) & Z^s_{nn}(\omega_{AC}) \end{bmatrix} \begin{bmatrix} \underline{I}_p(\omega_{AC}) \\ \underline{I}_n(\omega_{AC}) \end{bmatrix}, \text{ Equation (34)}$$

Fig. 3. Comparison of frequency-domain relations.

the $pn$-rotating formulation in (**4**), as shown in Fig. 1.

*A. Relation between pn-Stationary and pn-Rotating Frames.*

The transformation between the two $pn$-frames is a frequency shift, since $\omega_{DC} = \omega_{AC} - \omega_1$. For a symmetric impedance,

$$Z^r_{pp}(\omega_{DC}) = Z^s_{pp}(\omega_{AC} = \omega_{DC} + \omega_1)$$
$$\begin{cases} Z^r_{nn}(\omega_{DC}) = \left(Z^s_{pp}(\omega_{AC} = |\omega_{DC} - \omega_1|)\right)^* ; \omega_{DC} < \omega_1 \\ Z^r_{nn}(\omega_{DC}) = Z^s_{nn}(\omega_{AC} = \omega_{DC} + \omega_1) ; \omega_{DC} > \omega_1 \end{cases}, \quad (41)$$

Details are given in [12]. Similar expressions can be derived for the terms of the admittance matrix.

*B. Numerical examples.*

To illustrate the behavior for angular frequencies below $\omega_1$, known as the mirror-frequency effect, numerical examples are presented. For simplicity, only the frequency $f_{DC}$ is specified, with $\omega_{DC} = 2\pi.f_{DC}$. Since these properties hold for both impedance and admittance matrices, the latter will be used in the following discussion. Substituting $f_{DC} = 51$ Hz into the admittance-matrix form of Eq. (30) yields

$$\begin{bmatrix} \underline{I}_p(101) \\ \underline{I}_n(1) \end{bmatrix} = \begin{bmatrix} Y^r_{pp}(51) & Y^r_{pn}(51) \\ Y^r_{np}(51) & Y^r_{nn}(51) \end{bmatrix} \begin{bmatrix} \underline{U}_p(101) \\ \underline{U}_n(1) \end{bmatrix}. \quad (42)$$

In this case, the notation is unambiguous and clear, since $f_{DC} > f_1 = 50$ Hz. For $f_{DC} = 1$ Hz, the notation becomes more complicated due to the appearance of negative frequencies, and we obtain

$$\begin{bmatrix} \underline{I}_p(51) \\ \underline{I}_n(-49) \end{bmatrix} = \begin{bmatrix} \underline{I}_p(51) \\ \underline{I}_p(49) \end{bmatrix} = \begin{bmatrix} Y^r_{pp}(1) & Y^r_{pn}(1) \\ Y^r_{np}(1) & Y^r_{nn}(1) \end{bmatrix} \begin{bmatrix} \underline{U}_p(51) \\ \underline{U}_p(49) \end{bmatrix}. \quad (43)$$

Here, the fact that negative frequencies in the negative sequence are equivalent to positive frequencies in the positive sequence has been used, that is, $\underline{I}_n(f_{DC} - f_1) = \underline{I}_p(|f_{DC} - f_1|)$ and $\underline{U}_n(f_{DC} - f_1) = \underline{U}_p(|f_{DC} - f_1|)$. Another example involving negative frequencies is the case of $f_{DC} = 49$ Hz, from which we obtain

$$\begin{bmatrix} \underline{I}_p(99) \\ \underline{I}_n(-1) \end{bmatrix} = \begin{bmatrix} \underline{I}_p(99) \\ \underline{I}_p(1) \end{bmatrix} = \begin{bmatrix} Y^r_{pp}(49) & Y^r_{pn}(49) \\ Y^r_{np}(49) & Y^r_{nn}(49) \end{bmatrix} \begin{bmatrix} \underline{U}_p(99) \\ \underline{U}_p(1) \end{bmatrix}. \quad (44)$$

For the unified impedance model of Eq. (38), the admittance matrix for the case $f_{AC} = 1$ Hz, is given by

$$\begin{bmatrix} \underline{I}_p(1) \\ \underline{I}_n(-99) \end{bmatrix} = \begin{bmatrix} \underline{I}_p(1) \\ \underline{I}_p(99) \end{bmatrix} = \begin{bmatrix} Y^r_{pp}(-49) & Y^r_{pn}(-49) \\ Y^r_{np}(-49) & Y^r_{nn}(-49) \end{bmatrix} \begin{bmatrix} \underline{U}_p(1) \\ \underline{U}_p(99) \end{bmatrix}. \quad (45)$$

By comparing Eqs. (44) and (45), it can be concluded that Eqs. (30) and (38) are equivalent and describe the same physical relationship between the frequencies.

## VII. Discussion

Matrix transformations such as the Park transformation have been used in electrical machines for more than a century. In contrast, their application to VSC converters is much more recent. As a result, no common notation has yet been accepted by the research community, leading to several inconsistencies in notation.

*A. Notation problems.*

Some references define the terms of the $pn$-sequence domain admittance matrix using ambiguous notation. For instance, Eq. (46) corresponds to the admittance matrix in Eq. (2) of [23]:

$$\mathbf{Y}_{inv} = \begin{bmatrix} Y_p(\omega_p) & J_n(2\omega_0 - \omega_p) \\ J_p(\omega_p) & Y_n(2\omega_0 - \omega_p) \end{bmatrix}, \quad (46)$$

where the notation is unclear. In particular, it is difficult to determine how the equivalent impedances of series or parallel connections depend on $\omega_p$, or how $Z = Y^{-1}$ depends on $\omega_p$. A notation similar to Eq. (46) is also used in [24]. As shown in Fig. 3 of [25], the system is represented by the following matrix in the stationary reference frame with the following form

$$\begin{bmatrix} f_{pp}(s - j\omega_1) & f_{pn}(s + j\omega_1) \\ f_{np}(s - j\omega_1) & f_{nn}(s + j\omega_1) \end{bmatrix}, \quad (47)$$

where the "+" signs are incorrect. Eq. (48) corresponds to Eq. (11) of [26], in which the "+" sign in $(s + j\omega_1)$ is incorrect.

$$\begin{bmatrix} \Delta u_{\boldsymbol{\alpha\beta}} \\ e^{j2\omega t}\Delta u^*_{\boldsymbol{\alpha\beta}} \end{bmatrix} = \begin{bmatrix} Z_{dq,pp}(s - j\omega_1) & Z_{dq,pn}(s - j\omega_1) \\ Z_{dq,np}(s + j\omega_1) & Z_{dq,nn}(s + j\omega_1) \end{bmatrix} \begin{bmatrix} \Delta i_{\boldsymbol{\alpha\beta}} \\ e^{j2\omega t}\Delta i^*_{\boldsymbol{\alpha\beta}} \end{bmatrix}. \quad (48)$$

Similar notation issues are also present in [27]. In other cases, such as Eq. (49), which corresponds to Eq. (9) in [28], we have

$$\begin{bmatrix} I_p(f_p) \\ I_{p2}(f_p - 2f_1) \end{bmatrix} = \begin{bmatrix} Y_{11}(s) & Y_{12}(s) \\ Y_{21}(s) & Y_{22}(s) \end{bmatrix} \begin{bmatrix} U_p(f_p) \\ U_{p2}(f_p - 2f_1) \end{bmatrix}, \quad (49)$$

where the value of $s$ in the matrix $Y(s)$ should be $s - j2\pi f_1$. The correct notation is given in Eq. (38), which is also used in [29].

A different notation problem appears in [20]. In that work, a figure is presented showing the relationships among different reference frames, similar to the one presented in [3]. The notation problem is that the pn-frame used in [3] and [20] actually corresponds to the $\alpha\beta$-complex frame shown in Fig. 1.

*B. Comments about a notation standar.*

To correctly interpret the notation in the $pn$ reference frames, it is important to examine the example of a series impedance given in Eq. (18). In the $pn$-rotating frame (modified sequence-domain impedance), the voltage–current relationship using the admittance matrix is given by

$$\begin{bmatrix} I_p(s+j\omega_1) \\ I_n(s-j\omega_1) \end{bmatrix} = \begin{bmatrix} \frac{1}{R+(s+j\omega_1)L} & 0 \\ 0 & \frac{1}{R+(s-j\omega_1)L} \end{bmatrix} \begin{bmatrix} U_p(s+j\omega_1) \\ U_n(s-j\omega_1) \end{bmatrix}, \quad (50)$$

assuming the substitution $s \rightarrow j\omega_{DC}$. The coexistence of two different frequencies is captured through the notation $\boldsymbol{Z}^{+}(s)=R+(s+j\omega_1)L$ and $(\boldsymbol{Z}^{+}(s))^{*}=R+(s-j\omega_1)L$.

The previous examples illustrate the need for a standardized notation that is widely recognized. In the *dq*-real reference frame, which is the most commonly used notation in the literature, there are no significant issues. However, in the *dq*-complex and *pn*-rotating reference frames, notation ambiguities do arise, as shown previously. Regarding the notation in the sequence domain, it is important to emphasize that:

*1) Notation in Laplace domain*: For the theoretical development of the model equations in the Laplace domain, the *dq*-complex reference frame notation is a good choice. Examples of the use of the *dq*-complex notation can be found in [11], [19], and [30]. This notation provides a compact and elegant framework for analyzing complex systems, such as AC/DC networks, as demonstrated in [19]. The use of the universal impedance model is more difficult to generalize due to the presence of the term $e^{j2\omega t}$. Since it is equivalent to the *pn*-rotating reference frame in the frequency domain, it introduces unnecessary complexity into the notation.

*2) Notation in frequency domain*: The modified sequence-domain impedance is expressed in the frequency domain and is equivalent to the *pn*-rotating frame approach. Both formulations use the angular frequency $\omega_{DC}$. The universal impedance model is also equivalent to the previously presented formulations, however, it uses the angular frequency $\omega_{AC}$. One drawback of using $f_{AC}$ is that it introduces additional negative terms, as illustrated in Eq. (45).

## VIII. Conclusions

This paper reviews the transformations commonly used in impedance/admittance-based modeling of VSC systems and clarifies their relationship with sequence-domain representations. The relationships between the *dq*-real, *dq*-complex, *αβ*-real, *αβ*-complex, and *pn* domains were systematically analyzed. The results show that impedance equations expressed in the *dq*-complex and *αβ*-complex domains lead to different formulations in the sequence domain. It has been demonstrated that the universal impedance model is, in reality, formulated in a rotating reference frame. A key finding is that transformations between stationary and rotating frames are physically meaningful only when the impedance or admittance matrix is symmetric. This limitation is often overlooked in the literature. The physical interpretation presented in this work highlights the importance of clearly distinguishing between stationary and rotating reference frames when analyzing converter dynamics.

## Appendix I: Power Scaling

The problem of scaling in matrix transformations and its influence on power is described in detail in [10]:

$$\begin{pmatrix} u_\alpha \\ u_\beta \end{pmatrix} = T_{32} \begin{pmatrix} u_a \\ u_b \\ u_c \end{pmatrix} \quad ; \quad T_{32} = K \begin{bmatrix} \frac{2}{3} & -\frac{1}{3} & -\frac{1}{3} \\ 0 & \frac{1}{\sqrt{3}} & -\frac{1}{\sqrt{3}} \end{bmatrix}. \quad (51)$$

The scaling constant can be chosen as $K = \left\{1, 1/\sqrt{2}, \sqrt{3/2}\right\}$, for peak value, root-mean-square value, or power-invariant scaling, respectively [10]. The matrix $T_{32}$ corresponds to *P(0)* (see Eq. (1)) in the power-invariant scaling case.

The power expression for different scaling constants is

$$P = \frac{3}{2K^2}\left(u_x i_x + u_y i_y\right), \quad (52)$$

where (*x, y*) can represent (*α, β*) in the *αβ*-frame, (*d, q*) in the *dq*-frame or (*p, n*) in the *pn*-frame.

## Appendix II: Ku transformation of voltages

Consider a three-phase voltage system containing positive- and negative-sequence components with angular frequency $\omega_x$, defined as

$$\begin{aligned} & u_a = U_a \cos(\omega_x t + \alpha_a) \quad ; \quad u_b = U_b \cos(\omega_x t + \alpha_b) \\ & u_c = U_c \cos(\omega_x t + \alpha_c). \end{aligned} \quad (53)$$

When this system is subjected to the *Ku* transformation (3),

$$\mathbf{U_{dq}} = \frac{1}{\sqrt{3}}\left(e^{-j\omega_1 t} u_a + a \cdot e^{-j\omega_1 t} u_b + a^2 \cdot e^{-j\omega_1 t} u_c\right). \quad (54)$$

The result is obtained by applying Euler's formula

$$\begin{aligned} \mathbf{U_{dq}} = \frac{e^{-j\omega_1 t}}{\sqrt{3}} \Big\{ & U_a\left(e^{j(\omega_x t+\alpha_a)} + e^{-j(\omega_x t+\alpha_a)}/2\right) \\ & + a \cdot U_b\left(e^{j(\omega_x t+\alpha_b)} + e^{-j(\omega_x t+\alpha_b)}/2\right) \\ & + a^2 \cdot U_c\left(e^{j(\omega_x t+\alpha_c)} + e^{-j(\omega_x t+\alpha_c)}/2\right), \end{aligned} \quad (55)$$

and grouping the resulting terms

$$\begin{aligned} \mathbf{U_{dq}} = \frac{1}{2\sqrt{3}} \Big\{ & \left(U_a e^{j\alpha_a} + a \cdot U_b e^{j\alpha_b} + a^2 \cdot U_b e^{j\alpha_c}\right) e^{j(\omega_x - \omega_1)t} \\ & + \left(U_a e^{-j\alpha_a} + a \cdot U_b e^{-j\alpha_b} + a^2 \cdot U_b e^{-j\alpha_c}\right) e^{-j(\omega_x + \omega_1)t} \Big\}, \end{aligned} \quad (56)$$

which leads to the final expression

$$\mathbf{U_{dq}} = \underline{U}_p e^{j(\omega_x - \omega_1)t} + \underline{U}_n^{*} e^{-j(\omega_x + \omega_1)t}. \quad (57)$$

In Eq. (57), $\underline{U}_p$ and $\underline{U}_n$ represent the positive- and negative-sequence components, respectively, up to a scaling factor. The quantities of $\underline{U}_p$ and $\underline{U}_n$ differ from the conventional sequence components only by a scaling factor, resulting in the following relationship:

$$\underline{U}_p = \sqrt{\frac{3}{2}}\underline{U}_p^{st} \quad ; \quad \underline{U}_n = \sqrt{\frac{3}{2}}\underline{U}_n^{st}. \tag{58}$$

where the superscript *st* denotes the standard notation. These scaling factors are omitted to simplify the notation.

## APPENDIX III: PHYSICAL INTERPRETATION

This section provides a physical interpretation of the interaction between voltages at different frequencies in rotating reference frames.

### A. Physical Interpretation of the Ku Transformation

Appendix II shows that applying the Ku transformation $K(\theta)$, with $\theta = \omega_1 t = 2\pi f_1 t$, to a three-phase voltage system containing positive- and negative-sequence components at angular frequency $\omega_x$, produces two three-phase voltage systems: a positive-sequence component at $\omega_x - \omega_1$ and a negative-sequence component at $\omega_x + \omega_1$.

In this work, the relevant case consists of two three-phase systems at angular frequencies $\omega_{DC}+\omega_1$ and $\omega_{DC}-\omega_1$.

$$\left.\begin{aligned} u_{ap} &= U_p \cos\left(\left(\omega_{DC}+\omega_1\right)t+\alpha_p\right) \\ u_{bp} &= U_p \cos\left(\left(\omega_{DC}+\omega_1\right)t-2\pi/3+\alpha_p\right) \\ u_{cp} &= U_p \cos\left(\left(\omega_{DC}+\omega_1\right)t+2\pi/3+\alpha_p\right) \end{aligned}\right\} \\ \left.\begin{aligned} u_{an} &= U_n \cos(\left(\omega_{DC}-\omega_1\right)t+\alpha_n) \\ u_{bn} &= U_n \cos\left(\left(\omega_{DC}-\omega_1\right)t+2\pi/3+\alpha_n\right) \\ u_{cn} &= U_n \cos\left(\left(\omega_{DC}-\omega_1\right)t-2\pi/3+\alpha_n\right) \end{aligned}\right\}. \tag{59}$$

The system at $\omega_{DC}+\omega_1$ is a pure positive-sequence system. Applying the Ku transformation yields a positive-sequence component at $\omega_{DC}$ and a negative-sequence component at $\omega_{DC}+2\omega_1$, which vanish because no negative-sequence component exists. Similarly, the system at $\omega_{DC}-\omega_1$ must be a purely negative-sequence system. The Ku transformation produces a negative-sequence component at $\omega_{DC}$ and a component at $\omega_{DC}-2\omega_1$, which is zero because no positive-sequence exists. Thus, the *dq*-complex voltage becomes

$$\mathbf{U}_{\mathrm{dq}} = \underline{U}_p e^{j\omega_{DC}t} + \underline{U}_n^* e^{-j\omega_{DC}t}, \tag{60}$$

where $\underline{U}_p = U_p\angle\alpha_p$, and $\underline{U}_n = U_n\angle\alpha_n$.

### B. Physical Interpretation of VSC Converter Voltages

Fig. 4 shows a VSC connecting the AC and DC sides, modeled by three controlled AC voltage sources and a controlled DC current source. In the *dq* reference frame

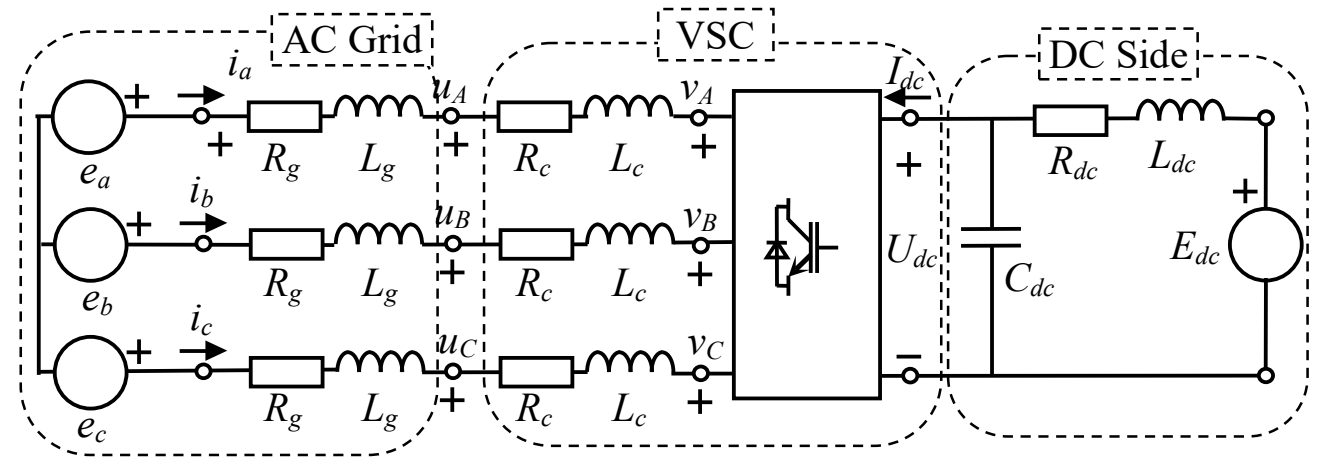

Fig. 4. Grid-connected VSC system.

$$\begin{pmatrix} u_d \\ u_q \end{pmatrix} = \begin{pmatrix} m_d \\ m_q \end{pmatrix} u_{DC}, \tag{61}$$

where $m_d$ and $m_q$ are modulation signals.

Applying the inverse Park transformation, $u_{abc} = P(\theta)^{\mathrm{T}} v_{dq}$, and $m_{abc} = P(\theta)^{\mathrm{T}} m_{dq}$., with $\theta = \omega_1 t$, reveals the interaction between the DC-side frequency $f_{DC}$ and the AC-side frequencies $f_p = f_{DC} + f_1$ and $f_n = f_{DC} - f_1$. Using Eq. (61), the phase voltages $u_{abc}$ are

$$\begin{aligned} u_a &= \left(M_{ac}\cos(\omega_1 t+\alpha_m)\right)u_{DC} \\ u_b &= \left(M_{ac}\cos(\omega_1 t+\alpha_m-2\pi/3)\right)u_{DC} \\ u_c &= \left(M_{ac}\cos(\omega_1 t+\alpha_m+2\pi/3)\right)u_{DC}, \end{aligned} \tag{62}$$

where $M_{ac}$ and $\alpha_m$ denote the modulation magnitude and phase in the *dq*-frame, and $u_{DC}$ is a DC-side perturbation voltage

$$u_{DC} = U_{DC}\cos(\omega_{DC}t+\alpha_{DC}), \tag{63}$$

with angular frequency $\omega_{DC} = 2\pi f_{DC}$. Multiplying Eq. (62) and (63) yields cosine products,

$$\begin{aligned} u_a &= \begin{pmatrix} M_{ac}U_{kDC}\frac{1}{2}\left\{\cos\left(\left[\omega_{DC}-\omega_1\right]t+\alpha_{DC}-\alpha_m\right)\right. \\ \left.+\cos\left(\left[\omega_{DC}+\omega_1\right]t+\alpha_{DC}+\alpha_m\right)\right\} \end{pmatrix} \\ u_b &= \begin{pmatrix} M_{ac}U_{kDC}\frac{1}{2}\left\{\cos\left(\left[\omega_{DC}-\omega_1\right]t+\alpha_{DC}-\alpha_m+2\pi/3\right)\right. \\ \left.+\cos\left(\left[\omega_{DC}+\omega_1\right]t+\alpha_{DC}+\alpha_m-2\pi/3\right)\right\} \end{pmatrix} \\ u_c &= \begin{pmatrix} M_{ac}U_{kDC}\frac{1}{2}\left\{\cos\left(\left[\omega_{DC}-\omega_1\right]t+\alpha_{DC}-\alpha_m-2\pi/3\right)\right. \\ \left.+\cos\left(\left[\omega_{DC}+\omega_1\right]t+\alpha_{DC}+\alpha_m+2\pi/3\right)\right\} \end{pmatrix}, \end{aligned} \tag{64}$$

which contains positive- and negative-sequence components at $\omega_p = \omega_{DC} + \omega_1$ and $\omega_n = \omega_{DC} - \omega_1$. After applying the Park transformation, the positive-sequence components produce terms at $\omega_{DC}$ and $\omega_{DC} + 2\omega_1$; while the negative-sequence components produce terms at $\omega_{DC}$ and $\omega_{DC} - 2\omega_1$. The terms $\omega_{DC} \pm 2\omega_1$ cancel out, leaving only components at $\omega_{DC}$.

Most studies assume an ideal DC source, in Fig. 4, and neglect DC-side voltage perturbations. However, PLLs and outer current controllers introduce asymmetry in the impedance matrix. Consequently, a positive-sequence current disturbance at $\omega_{DC}$ generates both positive- and negative-sequence voltage components, and the same occurs for a negative-sequence disturbance. When transformed back to the *abc*-frame, these components appear at $\omega_{DC}+\omega_1$ and $\omega_{DC}-\omega_1$. Therefore, in systems with an ideal DC-side, the disturbances also interact on the AC side at $\omega_{DC}\pm\omega_1$.

DC-side interactions can also be described using $3 \times 3$ impedance/admittance matrices [18], [19].

Eq. (65) represents the impedance matrix of a simplified AC-DC converter without control [12], in which the influence of $Z_{Tdc}(s) = R_{dc}+sL_{dc}$, see Fig. 4, is manifested through the nonzero off-diagonal terms,

$$\mathbf{Z}_{dq}^{+}(s) = \begin{bmatrix} \frac{3M^2}{4} Z_{Tdc}(s) + Z_c(s + j\omega_1) & \mathbf{m}_{\mathbf{dq}}^2 Z_{Tdc}(s) \\ (\mathbf{m}_{\mathbf{dq}}^*)^2 Z_{Tdc}(s) & \frac{3M^2}{4} Z_{Tdc}(s) + Z_c(s - j\omega_1) \end{bmatrix}. \quad (65)$$

Reference [12] provides a more detailed description of Eq. (65). It also includes the impedance matrix in the *dq*-real reference frame, as well as graphical representations of the terms $Y_{pp}(\omega)$, $Y_{pn}(\omega)$, $Y_{np}(\omega)$, and $Y_{nn}(\omega)$.

## REFERENCES

[1] J. Lesenne, F. Notelet, G. Seguier, “Introduction à l’Electrotechnique Approfondie”, Paris: Technique&Documentation, 1981.

[2] Y. H. Ku, "Rotating-Field Theory and General Analysis of Synchronous and Induction Machines," *Proceedings of the IEE* - Part IV: Institution Monographs, vol. 99, no. 4, pp. 410-428, December 1952.

[3] G.C. Paap, “Symmetrical components in the time domain and their application to power network calculations”, *IEEE Trans. Power Syst.*, vol. 15, no. 2, pp. 522-528, May 2000.

[4] M. Amin, M. Molinas, “Small-signal stability assessment of power electronics based power systems: A discussion of impedance- and eigenvalue-based methods,” *IEEE Trans. on Industry Applications,* vol. 53, no. 3, pp. 5014-5030, Sept/Oct. 2017.

[5] R.D. Middlebrook, “Input filter considerations in design and application of switching regulators“, in *Proc. IEEE Ind. Appl. Soc. Annu. Meeting*, Oct. 1976.

[6] B. Wen, D. Boroyevich, R. Burgos, P. Mattavelli and Z. Shen, "Analysis of D-Q Small-Signal Impedance of Grid-Tied Inverters," *IEEE Transactions on Power Electronics*, vol. 31, no. 1, pp. 675-687, Jan. 2016.

[7] M. Cespedes, J. Sun, “Impedance modeling and analysis of grid-connected voltage-source converters”, *IEEE Trans. Power Electron.*, vol. 29, no. 3, pp. 1254-1261, March 2014.

[8] A. Rygg, M. Molinas, C. Zhang, X. Cai, “On the equivalence and impact on stability of impedance modeling of power electronic converters in different domains”, *IEEE J. Emerg. Sel. Topics Power Electron.*, vol. 5, no. 4, pp. 1444-1454, Dec. 2017.

[9] A. Rygg, M. Molinas, C. Zhang, X. Cai, “A modified sequence-domain impedance definition and its equivalence to the dq-domain impedance definition for the stability analysis of AC power electronic systems,” *IEEE Journal of Emerging and Selected Topics in Power Electronics*, vol. 4, no. 4, pp. 1383-1396, Dec. 2016.

[10] L. Harnefors, “Modeling of Three-Phase Dynamic Systems Using Complex Transfer Functions and Transfer Matrices,“*IEEE Transactions on Industrial Electronics”*, vol. 54, no. 4, pp. 2239-2248, Aug. 2007.

[11] L. Harnefors, X. Wang, S. Chou, M. Bongiorno, M. Hinkkanen and M. Routimo, "Asymmetric Complex-Vector Models With Application to VSC–Grid Interaction," *IEEE Journal of Emerging and Selected Topics in Power Electronics*, vol. 8, no. 2, pp. 1911-1921, June 2020.

[12] J. Pedra, L. Sainz and L. Monjo, "Review and Improvements to the Measurements of the VSC Impedance Transfer Matrix,`" *IEEE Transactions on Power Delivery*, vol. 39, no. 2, pp. 1283-1298, April 2024.

[13] D. N. Zmood, D. G. Holmes and G. H. Bode, "Frequency-domain analysis of three-phase linear current regulators," *IEEE Transactions on Industry Applications*, vol. 37, no. 2, pp. 601-610, March-April 2001.

[14] B. Wen, D. Boroyevich, R. Burgos and P. Mattavelli, "Input impedance of voltage source converter with stationary frame linear current regulators and phase-locked loop," *2013 IEEE Energy Conversion Congress and Exposition*, Denver, CO, USA, 2013, pp. 4207-4213.

[15] J. Guo *et al.*, "Analysis and Mitigation of Low-Frequency Interactions Between the Source and Load Virtual Synchronous Machine in an Islanded Microgrid," *IEEE Transactions on Industrial Electronics*, vol. 69, no. 4, pp. 3732-3742, April 2022.

[16] X. Wang, L. Harnefors, F. Blaabjerg, “Unified impedance model of grid-connected voltage-source converters,” *IEEE Trans. on Power Electronics*, vol. 33, no. 2, pp. 1775-1785, Feb. 2018.

[17] Y. Liao and X. Wang, "Stationary-Frame Complex-Valued Frequency-Domain Modeling of Three-Phase Power Converters," *IEEE Journal of Emerging and Selected Topics in Power Electronics*, vol. 8, no. 2, pp. 1922-1933, June 2020.

[18] S. Shah, L. Parsa, "Impedance Modeling of Three-Phase Voltage Source Converters in DQ, Sequence, and Phasor Domains," *IEEE Transactions on Energy Conversion*, vol. 32, no. 3, pp. 1139-1150, Sept. 2017.

[19] J. Pedra, L. Sainz and L. Monjo, "Three-Port Small Signal Admittance-Based Model of VSCs for Studies of Multi-terminal HVDC Hybrid AC/DC Transmission Grids," *IEEE Transactions on Power Systems*, vol. 36, no. 1, pp. 732-743, Jan. 2021.

[20] G. Amico, A. Egea-Àlvarez, P. Brogan and S. Zhang, "Small-Signal Converter Admittance in the pn-Frame: Systematic Derivation and Analysis of the Cross-Coupling Terms," *IEEE Transactions on Energy Conversion*, vol. 34, no. 4, pp. 1829-1838, Dec. 2019.

[21] L. Meng, U. Karaagac, A. Pavani, J. Mahseredjian and K. Jacobs, "Examination of EMT-Type Impedance Scanning Techniques for Small Signal Stability Assessment of Inverter Based Resources," in *IEEE Transactions on Power Delivery*, vol. 40, no. 3, pp. 1607-1620, June 2025.

[22] L. Meng, U. Karaagac, and K. Jacobs, “A new sequence domain EMTlevel multi-input multi-output frequency scanning method for inverter based resources,” Elect. Power Syst. Res., vol. 220, 2023.

[23] X. Zou, X. Du, G. Wang, H. Tai, “ Single stationary domain equivalent inverter admittance for three-phase grid-inverter system considering the interaction between grid and inverter” *IET Power Electronics.*, Vol. 12, Iss. 6, 2019, pp. 1593-1602.

[24] M. K. Bakhshizadeh, X. Wang, F. Blaabjerg, J. Hjerrild, L. Kocewiak, C. L. Bak, B. Hesselbaek, “Couplings in phase domain impedance modeling of grid-connected converters”, *IEEE Trans. Power Electron*., vol. 31, no. 10, pp. 6792-6796, Oct. 2016.

[25] L. Chen, H. Nian, Y. Xu, “Complex transfer function-based sequence domain impedance model of doubly fed induction generator”, *IET Renew. Power Gener.*, Vol. 13, Iss. 1, 2019, pp. 67-77.

[26] X. Zhang, Y. Zhang, R. Fang and D. Xu, "Impedance Modeling and SSR Analysis of DFIG Using Complex Vector Theory," *IEEE Access*, vol. 7, pp. 155860-155870, 2019.

[27] S. Wang, H. Liu, J. Ma, L. Feng, T. Liu, Z. Wu, et al. “An accurate admittance model of grid-connected converter with droop control”, *Int J. Electr Power Energy Syst*. 2024;157:109856.

[28] Y. Xu, H. Nian, T. Wang, L. Chen and T. Zheng, "Frequency Coupling Characteristic Modeling and Stability Analysis of Doubly Fed Induction Generator, " *IEEE Transactions on Energy Conversion,* vol. 33, no. 3, pp. 1475-1486, Sept. 2018.

[29] X. Lin, H. Wen, J. Yu, J. Zhang and J. C. -H. Peng, "Role Determination of Impedance Coupling in GCC With DC-Link Virtual Inertia Control," *IEEE Transactions on Industrial Electronics*, vol. 71, no. 3, pp. 2533-2544, March 2024.

[30] J. Pedra, L. Sainz, Ll. Monjo, “Comparison of small-signal admittance based models of doubly-fed induction generators.” *Int J Electr Power Energy Syst.* 2023;145:108654.